\documentclass[12pt]{article}

\usepackage{graphicx}
\usepackage{amsmath,amssymb}
\usepackage{bm}
\usepackage{graphicx, color}
\usepackage{wrapfig}
\usepackage{epsf}
\usepackage{latexsym}

\begin{document}
\title{
\begin{flushright}
\ \\*[-80pt] 
\begin{minipage}{0.2\linewidth}
\normalsize
\end{minipage}
\end{flushright}
{\Large \bf Deviations from Tri-Bimaximality and \\ Quark-Lepton Complementarity}}
\lineskip .75em
\vskip 1.5cm

\author{
\centerline{
Yusuke~Shimizu$^{1,}$\footnote{E-mail address: shimizu@muse.sc.niigata-u.ac.jp} and  
Ryo Takahashi$^{2,}$\footnote{E-mail address: ryo.takahashi@mpi-hd.mpg.de}}
\\*[20pt]
\centerline{
\begin{minipage}{\linewidth}
\begin{center}
$^1${\it \normalsize
Graduate~School~of~Science~and~Technology,~Niigata~University, \\ 
Niigata~950-2181,~Japan } \\
$^2${\it \normalsize 
Max-Planck-Institut f$\ddot{u}$r Kernphysik, Saupfercheckweg 
1, \\ 
69117 Heidelberg, Germany } 
\end{center}
\end{minipage}}
\\*[50pt]}

\date{
\centerline{\small \bf Abstract}
\begin{minipage}{0.9\linewidth}
\medskip 
\medskip 
\small
We study deviations from tri-bimaximality (TBM) and quark-lepton 
complementarity (QLC) in a model independent way. The current neutrino 
experimental data is well approximated by tri-bimaximal generation mixing but 
the QLC relations are not satisfied with each data of 1$\sigma$ level. This 
means that there exist deviations from the complementarity. The same fact for 
the TBM might be checked in the future neutrino experiments. We discuss such 
deviations from the TBM and QLC, simultaneously. A new ratio between the 
deviations is introduced, and some interesting points are presented. We also 
show predicted correlations among leptonic mixing angles at the points.
\end{minipage}
}
\begin{titlepage}
\maketitle
\thispagestyle{empty}
\end{titlepage}

\section{Introduction}
The current precision measurements of neutrino oscillation have suggested 
that there are two large mixing angles among three generations in the lepton 
sector unlike the quark sector. It is known that the experimental data of 
mixing angles \cite{Schwetz:2008er} is approximated by the tri-bimaximal 
generation mixing \cite{TB}, which is described as
 \begin{eqnarray}
  V_{\text{TB}}=
   \left(
    \begin{array}{ccc}
     2/\sqrt{6}  & 1/\sqrt{3} & 0           \\
     -1/\sqrt{6} & 1/\sqrt{3} & -1/\sqrt{2} \\
     -1/\sqrt{6} & 1/\sqrt{3} & 1/\sqrt{2}
    \end{array}
   \right).
 \end{eqnarray} 
This matrix leads to the following values of mixing angles:
 \begin{eqnarray}
  \sin^2\theta_{12}^l=\frac{1}{3},~~~\sin^2\theta_{23}^l=\frac{1}{2},~~~
  \sin^2\theta_{13}^l=0,
 \end{eqnarray}
or equivalently,
 \begin{eqnarray}
  \theta_{12}^l\simeq35.3^\circ,~~~\theta_{23}^l=45^\circ,~~~\theta_{13}^l=0, 
  \label{TBM}
 \end{eqnarray}
where $\theta_{ij}^l$ $(i,j=1,2,3;~i<j)$ are the 
Pontecorvo-Maki-Nakagawa-Sakata (PMNS) \cite{Maki:1962mu} mixing angles. This 
is one of interesting theoretical suggestions, and thus, such a suggestive form
 of generation mixing matrix strongly motivates the search for a hidden flavour
 structure of the lepton sector. In fact, a number of proposals based on a 
flavour symmetry have been elaborated \cite{Altarelli:2005yp,A4}. The observed 
values of PMNS mixing angles from the current neutrino oscillation experiments 
\cite{Schwetz:2008er} are
 \begin{eqnarray}
  \theta_{12}^l = (34.3^{+1.16}_{-0.991})^\circ,~~~
  \theta_{23}^l = (45^{+4.02}_{-3.45})^\circ,~~~
  \theta_{13}^l = (6.55^{+2.73}_{-2.92})^\circ,   
 \end{eqnarray}
at 1$\sigma$ level. We find that there are small deviations of the best-fit 
values for solar and reactor angles from the TBM while the best-fit value of 
atmospheric angle equals the one of TBM.

Regarding generation mixing angles including the quark sector, 
intriguing relations among mixing angles of quark and lepton sectors have been 
proposed in \cite{Raidal:2004iw,Minakata:2004xt}, which is called quark-lepton 
complementarity (QLC).
The originally proposed QLC relation is described as 
 \begin{eqnarray}
  \theta_{12}^l+\theta_{12}^q=\frac{\pi}{4}=45^\circ, 
  \label{12}
 \end{eqnarray}
where $\theta_{ij}^q$ are the Cabibbo-Kobayashi-Maskawa mixing angles 
\cite{Cabibbo:1963yz}. The second and third QLC relations can also be written as
\begin{eqnarray}
  &&\theta_{23}^l+\theta_{23}^q=\frac{\pi}{4}=45^\circ, 
  \label{23} \\
  &&\theta_{13}^l+\theta_{13}^q=0. \label{13}
 \end{eqnarray}
In ref.~\cite{Raidal:2004iw}, a realization of QLC relations has been
 proposed in the context of the grand unified theory with non-Abelian flavour 
symmetry. Then implication of relations for the quark-lepton symmetry and the 
mechanism of neutrino mass generation has been discussed. 

The current mixing angles for the quark sector \cite{pdg} are given at 
1$\sigma$ as
 \begin{eqnarray}
  \sin\theta_{12}^q = 0.2257\pm0.0010,~~~
  \sin\theta_{23}^q = 0.0415^{+0.0010}_{-0.0011},~~~
  \sin\theta_{13}^q = 0.00359\pm0.00016,
 \end{eqnarray}
or equivalently,
 \begin{eqnarray}
  \theta_{12}^q = (13.0^{+0.118}_{-0.0588})^\circ,~~~
  \theta_{23}^q = (2.38^{+0.0573}_{-0.0631})^\circ,~~~
  \theta_{13}^q = (0.206^{+0.00917}_{-0.00917})^\circ.
 \end{eqnarray}
Therefore, we find from the current experimental data \cite{Schwetz:2008er,pdg}
 that the above relations \eqref{12} and \eqref{13} are not satisfied with each 
data of 1$\sigma$:
 \begin{eqnarray}
  \theta_{12}^l+\theta_{12}^q \simeq (47.4^{+1.21}_{-1.05})^\circ,~~~
  \theta_{13}^l+\theta_{13}^q \simeq (6.75^{+2.74}_{-2.93})^\circ.
 \end{eqnarray}
The second QLC relation \eqref{23} can be satisfied with each data of 
1$\sigma$ level:
 \begin{eqnarray}
  \theta_{23}^l+\theta_{23}^q &\simeq& (47.4^{+4.08}_{-3.51})^\circ.
 \end{eqnarray}

One of the most important missions for the neutrino oscillation experiments 
is to clarify whether the reactor angle is zero or not. The finiteness of the 
reactor angle means of course that the TBM is ruled out. The improvement of 
accuracy to determine the solar and atmospheric angles is also an important task. 
Since the QLC relations are related with the leptonic mixing angles, the 
correlations make us possible to investigate the TBM and QLC, simultaneously. 
In this letter, we focus on deviations from TBM and QLC, simultaneously, towards 
upcoming data from neutrino oscillation experiments. 

This paper is organized as follows. In the second section, we define deviations 
from the TBM and QLC, and discuss the relations among the leptonic mixing angles 
and magnitudes of deviations while focusing on the current experimental bounds 
and future sensitivity for measuring the mixing angle of the reactor neutrino. 
Next we introduce a new ratio between the deviations from TBM and QLC, and show 
a relation of leptonic mixing angles with the introduced ratio. We also point 
out four relatively interesting points of this ratio. Then correlations of 
three leptonic mixing angles are shown in these points. The third section is 
devoted to the summary of our results.

\section{Deviations from tri-bimaximality and quark-lepton complementarity}

\begin{figure}
\begin{minipage}[]{0.49\linewidth} 
\begin{center}
~~~~
(a)

~

\includegraphics[width=7.5cm]{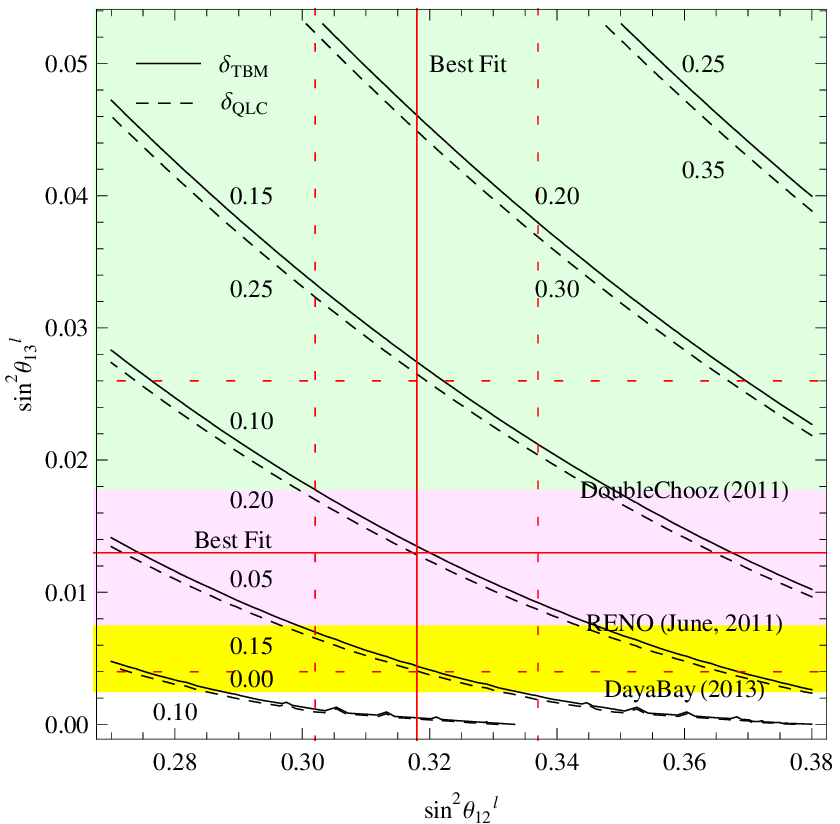}
\end{center}
\end{minipage}
\begin{minipage}[]{0.49\linewidth} 
\begin{center}
~~~~
(b)

~

\includegraphics[width=7.5cm]{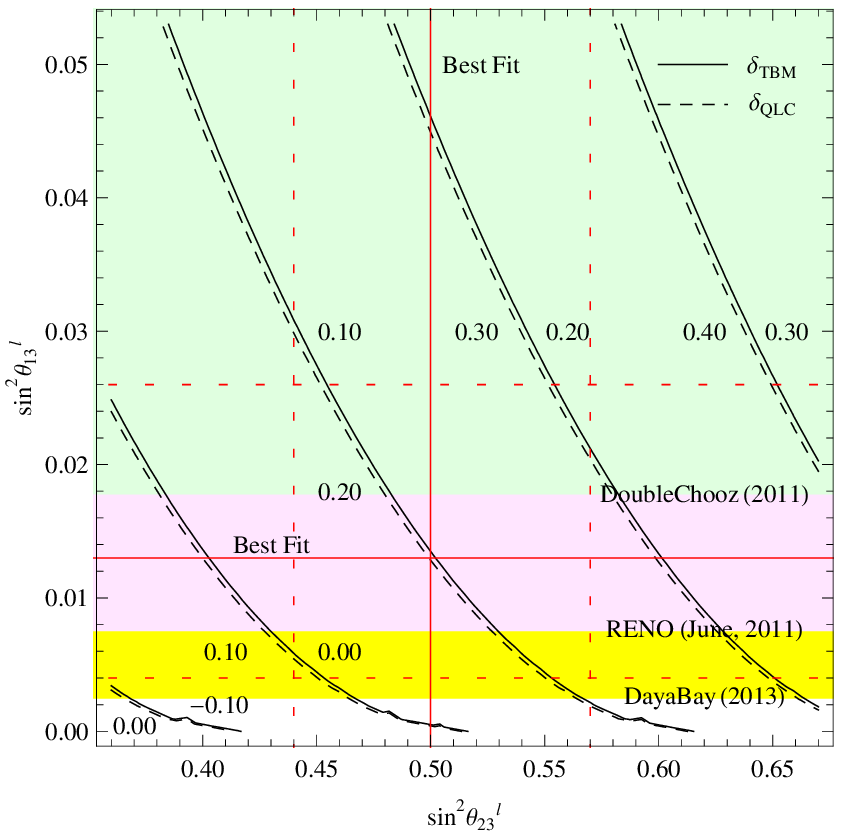}
\end{center}
\end{minipage}
\caption{Deviations from the TBM and QLC.}
\label{fig1}
\end{figure}

{}First, we define deviations from the TBM and QLC as
 \begin{eqnarray}
  \delta_{\text{TBM}}
   &\equiv& \sum_{i<j}
            [\theta_{ij}^l-\theta_{ij}^{\text{TBM}}], \label{dTBM} \\
  \delta_{\text{QLC}} 
   &\equiv& \sum_{i<j}
[(\theta_{ij}^l+\theta_{ij}^q)
                             -\theta_{ij}^{\text{QLC}}], \label{dQLC}
 \end{eqnarray}
respectively, where the mixing angles $\theta_{ij}^{l,q}$ are experimentally 
observed values, and 
$\theta_{12}^{\text{QLC}}=\theta_{23}^{\text{QLC}}=45^\circ$, 
$\theta_{13}^{\text{QLC}}=0$, and the values of $\theta_{ij}^{\text{TBM}}$ are 
given in \eqref{TBM}.\footnote{The discussions of experimentally 
observed values in the previous section are based on the PDG format of the 
Schechter-Valle parametrization \cite{pdg,Schechter:1980gr}. However, since the
 parameters indicating deviations \eqref{dTBM} and \eqref{dQLC} are defined by 
just mixing angles, the results given in this letter do not depend on the 
parametrization for the CKM and PMNS mixing matrices. We also discuss in 
ranges of $0\leq\theta_{ij}\leq\pi/2$ for all mixing angles throughout of this 
letter.} Figure \ref{fig1} shows the contour plots of these deviations in the 
plane of leptonic mixing angles. 
Each figure is drawn within 3$\sigma$ level for each mixing angle, and the 
contours are given in radian units. The solid and dashed lines correspond to 
the values of best fit and 1$\sigma$ range, respectively. The best-fit values 
of atmospheric and solar angles are utilized for the figs.~1(a) and (b), 
respectively, and the best-fit values of all CKM mixing angles are taken in 
both figures. The solid and dashed curves denote the deviations from the TBM 
and QLC defined in \eqref{dTBM} and \eqref{dQLC}, respectively. These figures 
can give clear comparisons and understandings for the deviations from the 
TBM and QLC. It is found that the magnitude of deviation from the QLC is larger
 than that from the TBM. For references, expected upper limits at 90\% CL for 
$\theta_{13}^l$, which are achieved by the DoubleChooz \cite{Ardellier:2006mn},
 RENO \cite{:2010vy}, and DayaBay \cite{Guo:2007ug} experiments one after 
another, are also shown by coloured regions. The times given in brackets for 
each experiment are roughly estimated by the values of 
$\sin^22\theta_{13}^l=0.07$, 0.03, and 0.01 for the DoubleChooz, RENO, and 
DayaBay experiments, respectively. Especially for the DayaBay experiment, the 
time is estimated from the expectation with the strongest sensitivity 
assumption.\footnote{See \cite{Mezzetto:2010zi} for a recent excellent review 
about the present status and prospect of $\theta_{13}^l$ measurements.}
 
\begin{figure}
\begin{minipage}[]{0.49\linewidth} 
\begin{center}
(a)

~

\includegraphics[width=7.5cm]{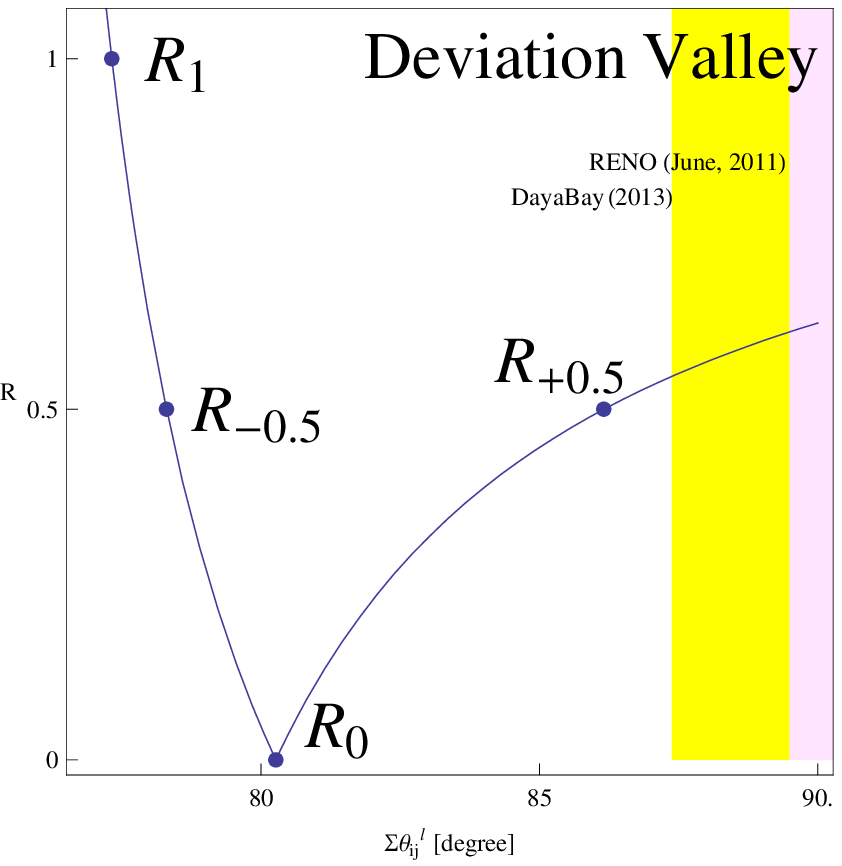}
\end{center}
\end{minipage}
\begin{minipage}[]{0.49\linewidth} 
\begin{center}
(b)

~

\includegraphics[width=7.5cm]{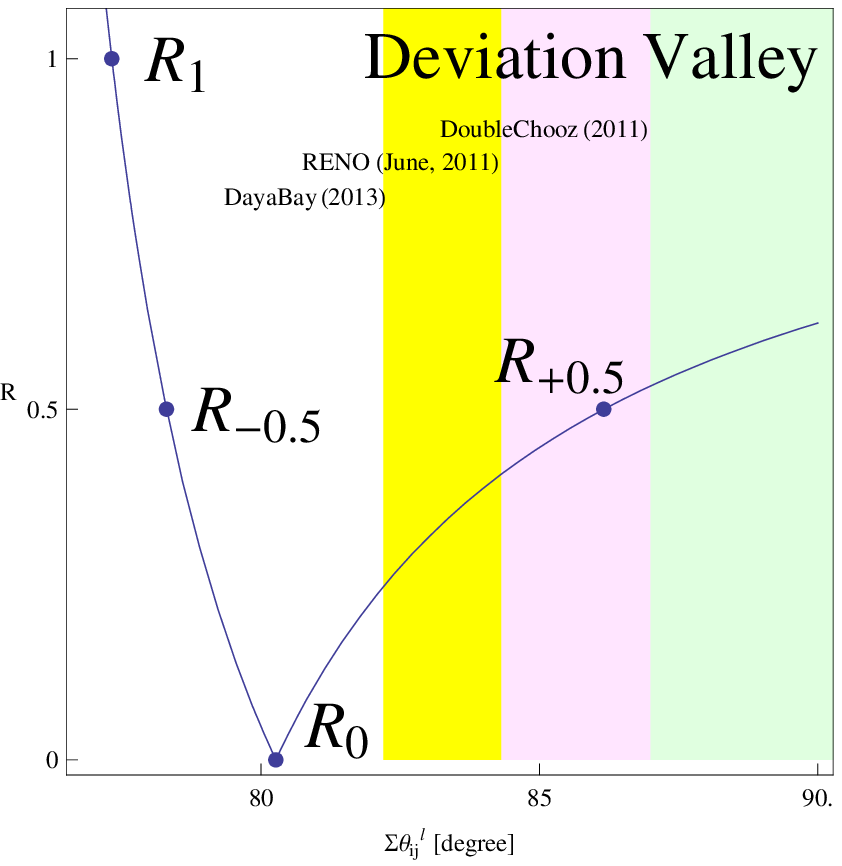}
\end{center}
\end{minipage}
\begin{center}
~~~~~~~~

(c)

~

\includegraphics[width=7.5cm]{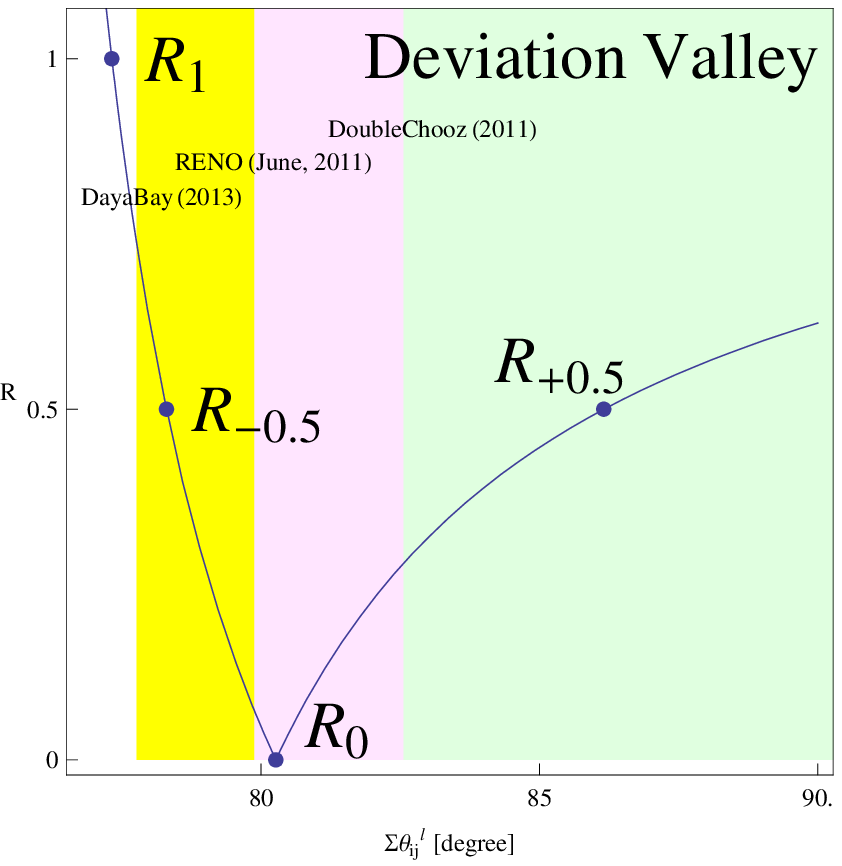}
\end{center}
\caption{Ratio between deviations from TBM and QLC.}
\label{fig2}
\end{figure}

Next, let us introduce a new ratio between the deviations from TBM and QLC 
towards a more profound understanding of deviations as
 \begin{eqnarray}
  R\equiv\frac{|\delta_{\text{TBM}}|}{\delta_{\text{QLC}}}.
 \end{eqnarray}
This ratio can take a finite positive value or zero because of the positive 
value of $\delta_{\text{QLC}}$ with each data of 1$\sigma$ level. The ratio 
becomes zero at the tri-bimaximal limit. Figure \ref{fig2} shows this ratio 
as a function of the sum of leptonic mixing angles, 
$\sum\theta_{ij}^l$.
The best-fit values of all CKM mixing angles have been taken. Since the 
deviation from the TBM, $\delta_{\text{TBM}}$, can become negative, the 
function of this ratio forms like a {\it valley}. The tri-bimaximal mixing 
corresponds to the bottom of the valley. We also show expected upper limits at 
90\% CL for $\theta_{13}^l$, which might be achieved by the DoubleChooz, RENO, 
and DayaBay experiments with the maximal values for both $\theta_{12}^l$ and 
$\theta_{23}^l$ at $1\sigma$ level in fig.~\ref{fig2}(a), with the best-fit 
values for the ones in fig.~\ref{fig2}(b), and with the minimal values in 
fig.~\ref{fig2}(c). These show relations among the ratio $R$ and the sum of the
 leptonic mixing angles, and future sensitivities for the measurement of 
$\theta_{13}^l$ in the upcoming neutrino oscillation experiments.

It might be the most suggestive case for particle physics if the exact tri-bimaximal
 mixing (bottom of the valley) could be satisfied in Nature. In this letter, we point
 out other suggestive scenarios in terms of deviations from the TBM and QLC. 
They are labelled by $R_1$, $R_{\pm0.5}$, and $R_0$ in fig.~\ref{fig2}. We 
call those points as {\it even}, {\it half}$_\pm$, and {\it cancelling 
deviation} scenarios, respectively, and the values of the ratio at these points are
 as follows:

{\it even deviation}
 \begin{eqnarray}
  R_1 &:& R=1~~~\hspace{0.5mm}\mbox{ and }~\delta_{\text{TBM}}=-\delta_{\text{QLC}}, 
 \end{eqnarray}

{\it half}$_-$ {\it deviation}
 \begin{eqnarray}
  R_{-0.5} &:& R=0.5~\mbox{ and }~\delta_{\text{TBM}}=-\frac{1}{2}\delta_{\text{QLC}}, 
 \end{eqnarray}

{\it half}$_+$ {\it deviation}
 \begin{eqnarray}
  R_{+0.5} &:& R=0.5~\mbox{ and }~\delta_{\text{TBM}}=+\frac{1}{2}\delta_{\text{QLC}}, 
 \end{eqnarray}

\begin{figure}
\begin{minipage}[]{0.4\linewidth} 
\begin{center}
{\it even}

~

\includegraphics[width=6.5cm]{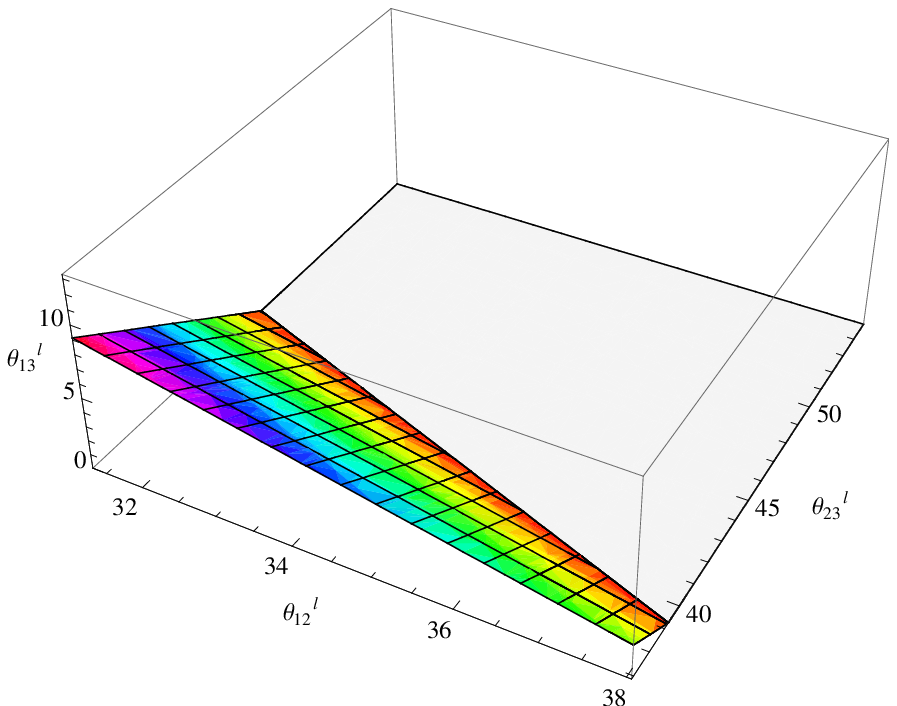}
\end{center}
\end{minipage}
\hspace{2cm}
\begin{minipage}[]{0.4\linewidth} 
\begin{center}
{\it half}$_-$

~

\includegraphics[width=6.5cm]{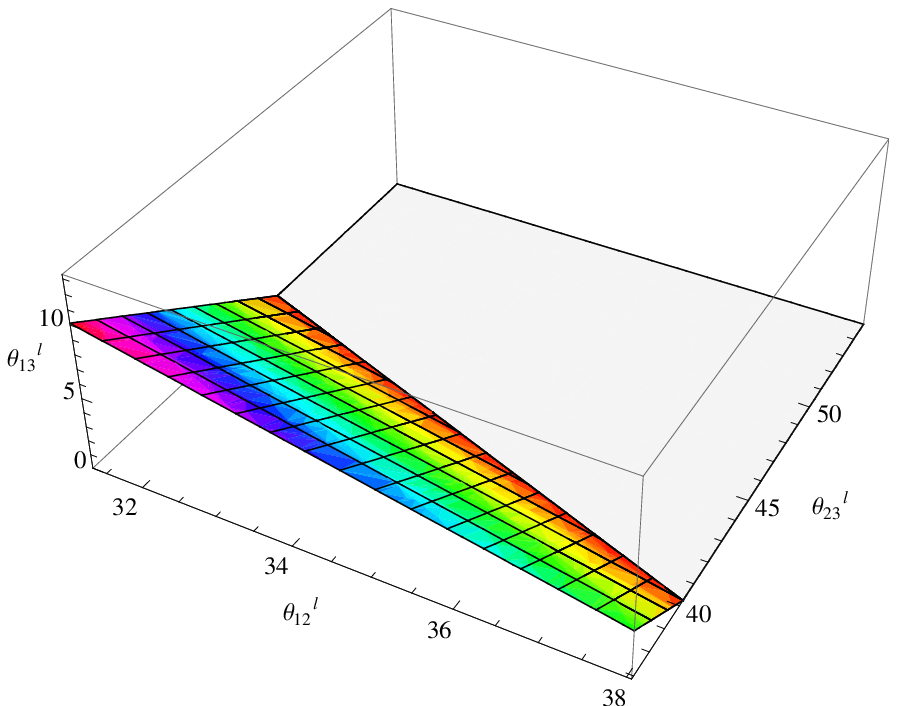}
\end{center}
\end{minipage}

\vspace{0.8cm}

\begin{minipage}[]{0.4\linewidth} 
\begin{center}
{\it half}$_+$

~

\includegraphics[width=6.5cm]{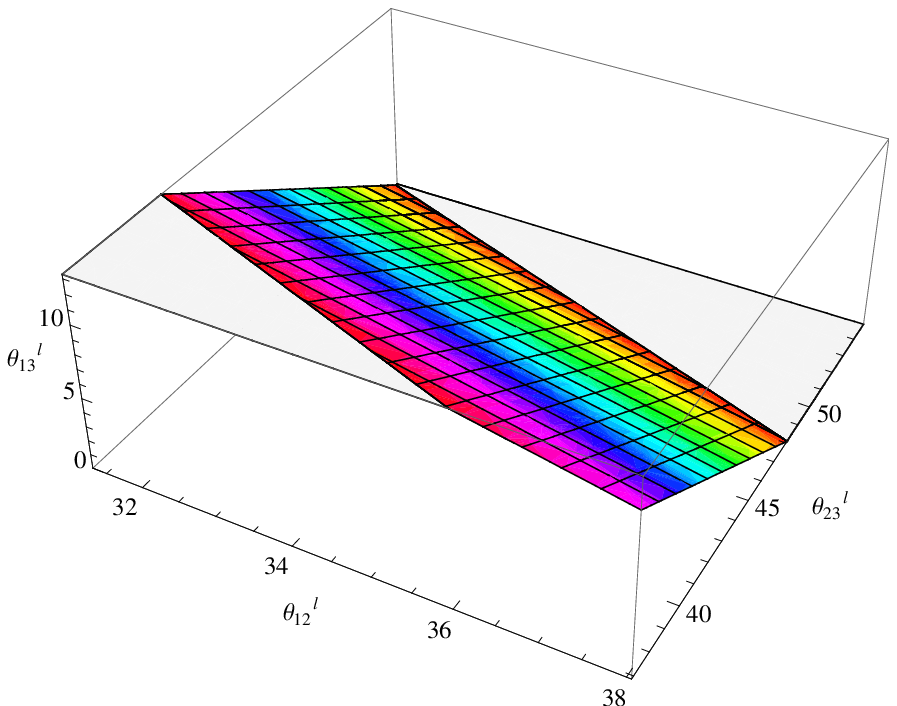}
\end{center}
\end{minipage}
\hspace{2cm}
\begin{minipage}[]{0.4\linewidth} 
\begin{center}
{\it cancelling}

~

\includegraphics[width=6.5cm]{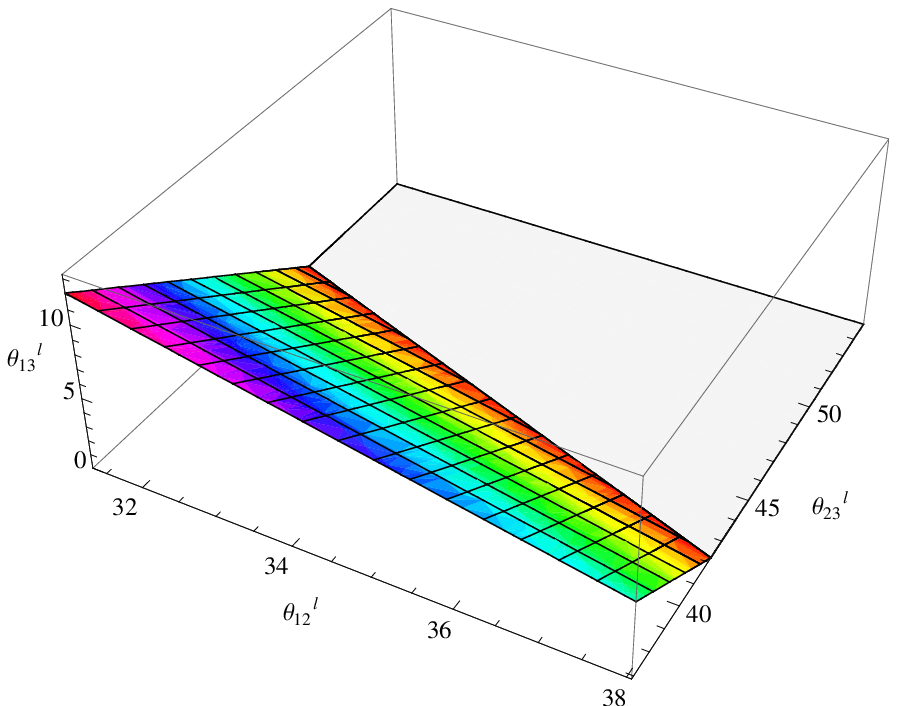}
\end{center}
\end{minipage}

\caption{Predicted surfaces of leptonic mixing angles from four scenarios.}
\label{fig3}
\end{figure}

{\it canceling deviation}
 \begin{eqnarray}
  R_0 &:& R=0~\mbox{ and }~\delta_{\text{TBM}}=0.  
 \end{eqnarray}
Notice that the exact tri-bimaximal mixing corresponds to the point labeled by
 $R_0$ but this point does not necessarily mean only the exact tri-bimaximal 
mixing, that is, that includes cancelled solutions among the deviation from TBM 
(see \eqref{dTBM} for the definition of deviation). The even deviation scenario 
shown by $R_1$ means that the absolute value of deviation from TBM equals 
that from the QLC, one can achieve this scenario with leptonic mixing angles at 
$3\sigma$ range. This point can be realized by a negative value of 
$\delta_{\text{TBM}}$. Finally, half$_\pm$ deviation scenarios shown by $R_{\pm0.5}$ 
denote that the magnitude of deviation from TBM becomes the half of that from 
QLC, whose point can be obtained from both positive and negative 
$\delta_{\text{TBM}}$. The half$_+$ and half$_-$ deviation scenarios are 
distinguished by the positive and negatives values of $\delta_{\text{TBM}}$, 
respectively. Notice that once the value of $R$ and the sign of 
$\delta_{\text{TBM}}$ are fixed we can find a unique surface in the space of 
leptonic mixing angles. The predicted surfaces corresponding to each deviation 
scenario are shown in fig.~\ref{fig3}. It is easily seen that the value of the 
reactor mixing angle for fixed values of atmospheric and solar mixing angles 
becomes larger as we proceed from $R_1$ to $R_{+0.5}$ through the valley.

All flavour models to discuss quark/lepton mixing are on the curve shown in 
fig.~\ref{fig2}. It is worth studying classification of all flavour models and 
constructing a model, which realizes suggestive scenarios based on this point 
of view about the ratio between deviations. In our new direction of 
simultaneous discussion about deviations from the TBM and QLC, it might be 
still very interesting if experimentally determined leptonic mixing angles 
could be somewhere in surfaces shown in fig.~\ref{fig3} except for the exact
 tri-bimaximal point, $\sin^2\theta_{12}=1/3$, $\sin^2\theta_{23}=1/2$, and 
$\sin\theta_{13}=0$. Without our new direction, one would argue about 
deviations from the TBM and QLC for each mixing angle, independently. However, 
once we introduce deviation parameters, $\delta_{\text{TBM}}$, 
$\delta_{\text{QLC}}$, and $R$, leptonic mixing angles somewhere in surfaces
 of fig.~\ref{fig3} can suggest {\it even}, {\it half}$_\pm$, or {\it canceling
 deviation} scenario. That will strongly motivate the construction of flavour 
models which should clarify a new physics (mechanism) behind the deviations 
from TBM and QLC in addition to investigation of the origin of TBM and QLC. For 
instance, if the observed solar and reactor angles are in the canceling scenario 
with maximal atmospheric angle, that means that the magnitude of deviation of 
the solar angle from the tri-bimaximal solar mixing is just the same as the size 
of the reactor angle;
 \begin{eqnarray}
  \sin^2(\theta_{12}^l+\theta_{13}^l)=\frac{1}{3}, \label{example}
 \end{eqnarray} 
when $\sin^2\theta_{23}^l=1/2$. We note that this example \eqref{example} shows
 a correlation between solar and reactor angles, which includes the exact 
tri-bimaximal mixing angles as the most suggestive point. Such relatively 
model-dependent studies based on our proposal and further discussions about 
resultant predictions will be presented in separate publications.

Finally, we show an estimation of the ratio $R$ in the $A_4$ model 
\cite{A4} as an example. In ref. \cite{A4}, the ratio is calculated as $R=0$ 
since the model can predict the exact tri-bimaximal mixing at the leading 
order. However, the next-to-leading order (NLO) corrections give deviations 
from the TBM \cite{Hayakawa}. Typical values of leptonic mixing angles up to 
the NLO are estimated as 
\begin{align}
\sin ^2\theta _{12}=0.36,~~~
\sin ^2\theta _{13}=4.8\times 10^{-6},~~~
\sin ^2\theta _{23}=0.48.
\end{align}
Therefore, the value of the ratio becomes $R\simeq0.085\simeq\mathcal{O}(0.1)$,
 where the best-fit values of all CKM mixing angles are taken.

\section{Summary}

In this letter, we have studied deviations from TBM and QLC in a model-independent 
way. First, we have defined those deviations, and then, presented those contours 
while comparing with upcoming reactor neutrino experiments. Once we fixed the best-fit 
value of the solar or atmospheric angle, the deviation from QLC is larger than that 
from TBM. Next, a new ratio between deviations from TBM and QLC has been introduced. 
We have focused on the ratio, and pointed out relatively suggestive four scenarios, 
which were named as even, half$_\pm$ and canceling deviation scenarios. Each scenario 
can predict a different surface with a correlation among the three leptonic mixing 
angles. If the future neutrino oscillation experiments could suggest that observed 
values of mixing angles are somewhere in the above scenario, our new proposal to 
understand deviations from TBM and QLC would strongly motivate the search for further 
hidden flavour structures of the standard model in addition to a clarification of
 the origins of TBM and QLC. It might be worth recognizing that some exotic 
correlations among mixing angles can be predicted in relatively model-dependent
 ways.

\subsection*{Acknowledgements}
We would like to thank M. Lindner and M. Tanimoto for helpful comments. The 
work of Y.S. is supported by the Japan Society of Promotion of Science. The 
work of R.T. is supported by the DFG-SFB TR 27. Y.S. is grateful to 
Max-Planck-Institut f$\ddot{\mbox{u}}$r Kernphysik for their hospitality.

\end{document}